\documentclass[aps,prd,preprintnumbers,amsmath,amssymb,nofootinbib,11pt]{revtex4}
\usepackage{eepic}
\usepackage{indentfirst}
\usepackage{mathrsfs}
\usepackage{fancyhdr}
\usepackage{ulem}
\usepackage{float}
\usepackage{graphicx}
\usepackage[
colorlinks=true,
linkcolor=black,
breaklinks=true,
urlcolor=blue,
citecolor=green]{hyperref}
\usepackage{epstopdf}

\newcommand{\be}{\begin{equation}}
\newcommand{\ee}{\end{equation}}
\newcommand{\ba}{\begin{array}{c}}
\newcommand{\ea}{\end{array}}
\renewcommand{\L}{\mathscr{L}}
\newcommand{\M}{\mathscr{M}}

\newcommand{\bra}{\langle}
\newcommand{\ket}{\rangle}
\newcommand{\nn}{\nonumber}
\newcommand{\MeV}{\,\text{MeV}}
\newcommand{\GeV}{\,\text{GeV}}
\renewcommand{\vec}[1]{\mathbf{#1}}
\newcommand{\diff }{{\text{d}}}

\begin{document}

\title{\boldmath Effect of $Z_b$ states on $\Upsilon(3S)\to\Upsilon(1S)\pi\pi$ decays}

\author{ Yun-Hua~Chen$^a$}\email{chen@hiskp.uni-bonn.de}
\author{ Johanna T.\ Daub$^{a}$}
\author{ Feng-Kun Guo$^{b,a}$}
\author{ Bastian Kubis$^{a}$}
\author{ Ulf-G.\ Mei\ss ner$^{a,c}$}
\author{ Bing-Song~Zou$^{b}$}
\affiliation{ ${}^a$Helmholtz-Institut f\"ur Strahlen- und Kernphysik (Theorie) and\\
             Bethe Center for Theoretical Physics,
             Universit\"at Bonn,
             53115 Bonn, Germany\\
             ${}^b$State Key Laboratory of Theoretical Physics,
             Institute of Theoretical Physics, CAS, Beijing 100190, China \\
             ${}^c$Institute for Advanced Simulation and
             J\"ulich Center for Hadron Physics,
             Institut f\"ur Kernphysik,
             Forschungszentrum J\"ulich,
             52425 J\"ulich, Germany
}

\begin{abstract}

Within the framework of dispersion theory, we analyze the dipion
transitions between the lightest $\Upsilon$ states, $\Upsilon(nS)
\rightarrow \Upsilon(mS) \pi\pi$ with $m < n \leq 3$. In particular,
we consider the possible effects of  two intermediate
bottomoniumlike exotic states $Z_b(10610)$ and $Z_b(10650)$. The
$\pi\pi$ rescattering effects are taken into account in a
model-independent way using dispersion theory. We confirm that
matching the dispersive representation to the leading chiral
amplitude alone cannot reproduce the peculiar two-peak $\pi\pi$ mass
spectrum of the decay $\Upsilon(3S) \rightarrow \Upsilon(1S)
\pi\pi$. The existence of the bottomoniumlike $Z_b$ states can
naturally explain this anomaly. We also point out the necessity of a
proper extraction of the coupling strengths for the $Z_b$ states to
$\Upsilon(nS)\pi$, which is only possible if a Flatt\'e-like
parametrization is used in the data analysis for the $Z_b$ states.

\end{abstract}


\maketitle

\newpage

\section{Introduction}

The hadronic transitions $\Upsilon(nS) \to \Upsilon(mS) \pi \pi $
between $\Upsilon$ states of different radial excitation numbers
$n$, $m$ are important processes for the understanding of systems
with both  heavy-quarkonium dynamics and low-energy QCD. Because of
the large $b$ quark mass, bottomonia as nonrelativistic $b\bar b$
bound states are expected to be compact. The light hadrons such as
pions emitted in the transitions between two bottomonia are normally
expected to be due to the hadronization of soft gluons. Thus, the
method of QCD multipole expansion together with soft-pion
theorems~\cite{Voloshin1980,Novikov1981,Kuang1981,Kuang2006} is
often used to study these transitions. This means that such a method
can be used to describe transitions where nonmultipole effects, such
as coupled-channel effects and intermediate resonances, are small
and the pions are very soft, such that the $\pi\pi$ final-state
interaction (FSI) can be neglected. A characteristic of this method
explored by the Cornell~\cite{Cornell1978,Cornell1980,TMYan1980} and
Orsay~\cite{Orsay1973,Orsay1977:1,Orsay1977:2} groups is that the
decay amplitudes are oscillatory functions of the decay momentum,
which is a direct consequence of the radial node structure in the
parent quarkonia wave functions. This can explain the ratio of
partial decay widths
$\Gamma(\Upsilon(3S)\to\Upsilon(1S)\pi\pi)/\Gamma(\Upsilon(2S)\to\Upsilon(1S)\pi\pi)\simeq
0.16$, though the phase space in the
$\Upsilon(3S)\to\Upsilon(1S)\pi\pi$ process is much larger than that
in $\Upsilon(2S)\to\Upsilon(1S)\pi\pi$, instead of interpretations
of the initial quarkonia states as $B^{(*)}$--$\bar{B}^{(*)}$
molecules as in Ref.~\cite{Glashow}. The $\pi\pi$ mass spectra of
the transitions between $2S$ and $1S$ heavy quarkonia can also be
well described by such a method.\footnote{The dipion invariant mass
distributions for both $\Upsilon(2S)\to\Upsilon(1S)\pi\pi$ and
$\psi(2S)\to J/\psi\pi\pi$ can be well described regardless of
whether the $\pi\pi$ FSI is included; see Ref.~\cite{Guo:2006ya}.
This is due to the simple shape of the $\pi\pi$ invariant mass
distributions in these cases and does not mean that the FSI is not
important. We also want to point out that the formula derived from
the QCD multiple expansion together with the soft-pion theorem was
used very often by experimentalists to fit their excellent data on
the dipion transitions between various heavy quarkonia. However,
this is often unjustified since the pions in these transitions are
not always soft. A good example is the decay
$\Upsilon(4S)\to\Upsilon(1S)\pi\pi$, where the dipion invariant mass
can take values of more than $1\GeV$, so that the FSI should not be
neglected~\cite{Guo:2006ai}.} However, there has been a well-known
anomaly for the dipion transitions: the data for the decay
$\Upsilon(3S) \to \Upsilon(1S) \pi \pi $ has a two-hump behavior,
while a naive application of the formula~\cite{Brown:1975dz} that
worked well for the $2S\to 1S$ and $3S\to 2S$ transitions would only
give a single peak at large dipion invariant masses. Many mechanisms
have been studied to explain this discrepancy, such as (i)
coupled-channel effects with open-bottom intermediate
states~\cite{Lipkin1988,Kuang1991,Simonov2009}, (ii) the existence
of a hypothetical resonance which couples to
$\Upsilon\pi$~\cite{Voloshin1983,Zou1995,Guo2005}, (iii) the
$\pi\pi$ resonance [the $f_0(500)$ or $\sigma$ meson] or strong
$\pi\pi$ final-state
interaction~\cite{Komada1,Komada2,Uehara,Moxhay1989,Chakravarty1994,Guo2005,Surovtsev2015},
(iv) relativistic corrections~\cite{Voloshin2006}, etc. Among these
mechanisms, the hypothetical $\Upsilon\pi$ resonance is in fact a
tetraquark state with quark content $b\bar b q\bar q$ and quantum
numbers $I^G(J^P)=1^+(1^+)$. The discovery of two $Z_b$ resonances
in channels including both $\Upsilon(1S)\pi$ and $\Upsilon(3S)\pi$
by the Belle Collaboration in 2011~\cite{Belle2011:1,Belle2012:1}
with such quantum numbers necessitates a reanalysis of
$\Upsilon(3S)\to \Upsilon(1S)\pi\pi$, taking into account these
resonances with their measured properties. Furthermore, since the
dipion invariant mass reaches almost 900~MeV in such a decay, and
the $\pi\pi$ $S$-wave FSI is known to be strong in this energy
range, it is thus also necessary to account for the $\pi\pi$ FSI
properly. Therefore, in the present paper we will use a formalism
incorporating mechanisms (ii) and (iii) above, with (ii) upgraded to
include the $Z_b$ states with measured properties given in the next
paragraph, and (iii) such that the $\pi\pi$ FSI is treated in a
model-independent way consistent with the $\pi\pi$ scattering data.
The coupled-channel effects will be commented upon very briefly at
the end of the paper; since we will use the leading-order
heavy-quark expansion we will neglect any relativistic corrections.

The two charged bottomoniumlike resonances $Z_b(10610)^\pm$ and
$Z_b(10650)^\pm$ were observed in the decay processes $\Upsilon(5S)
\rightarrow \Upsilon(nS) \pi^+ \pi^- $ ($n=1, 2, 3$) and
$\Upsilon(5S) \rightarrow h_b(mP) \pi^+ \pi^- $ ($m=1,
2$)~\cite{Belle2011:1,Belle2012:1}. Their quantum numbers are indeed
$I^G(J^P)=1^+(1^+)$, and their masses and widths have been
determined to be $M(Z_b(10610)) = (10607.4\pm 2.0)\MeV$,
$\Gamma(Z_b(10610)) = (18.4\pm 2.4)\MeV$, and $M(Z_b(10650)) =
(10652.2\pm 1.5)\MeV$, $\Gamma(Z_b(10650)) = (11.5\pm 2.2)\MeV$,
respectively. Preliminary results for the branching fractions of
$Z_b(10610)$ and $Z_b(10650)$ decays into $\Upsilon(nS)\pi^+ $
($n=1, 2, 3$) were also reported~\cite{Belle2012:2}.

We will therefore study the decays $\Upsilon(nS) \rightarrow
\Upsilon(mS) \pi\pi$ ($m < n\leq 3$), considering effects of the
$Z_b$ states. We will use dispersion theory in the form of modified
Omn\`es solutions to take into account the $\pi\pi$ FSIs. Herein,
the $Z_b$-exchange amplitudes provide a left-hand-cut contribution
to the dispersion integral. With the constraints of unitarity and
analyticity, the decay amplitude is determined up to a few
subtraction functions, which can be matched to the leading chiral
tree-level amplitude in the low-energy region. We adopt the leading
chiral Lagrangian for the coupling of two $S$-wave heavy quarkonia
to an even number of pions from Ref.~\cite{Mannel}, constructed in
the spirit of chiral perturbation theory and the heavy-quark
nonrelativistic expansion. The theoretical framework is described in
detail in Sec.~\ref{theor}. In Sec.~\ref{pheno}, we fit the decay
amplitudes to the data for the dipion transitions between two
$\Upsilon(nS)$ states. Through fitting the experimental data of the
$\pi\pi$ invariant mass distribution and the pion helicity angular
distribution, the low-energy constants (LECs) in the chiral
Lagrangian and the product of couplings for $Z_b \Upsilon\pi$ and $
Z_b \Upsilon^{\prime}\pi$ [here we use $\Upsilon$ and $\Upsilon'$ to
refer to the $\Upsilon(nS)$ in the final and initial states,
respectively] are determined. A brief summary and discussion will be
presented in Sec.~\ref{conclu}. Some details related to the matching
of the dispersive representation as well as the Flatt\'e
parametrization are relegated to Appendixes~\ref{Appendix.A}
and~\ref{Appendix.B}, respectively.

\section{Theoretical framework}
\label{theor}

\subsection{ Tree-level amplitudes}
\label{sect.lag}

The decay amplitude for
\be
\Upsilon(nS)(p_a) \to \Upsilon(mS)(p_b) \pi(p_c)\pi(p_d)
\ee
is described in terms of the Mandelstam variables
\begin{align}
s &= (p_c+p_d)^2 , \qquad
t=(p_a-p_c)^2\,, \qquad u=(p_a-p_d)^2\,,\nn\\
3s_0&\equiv
s+t+u=
 m_{\Upsilon(nS)}^2+m_{\Upsilon(mS)}^2+2m_\pi^2  \,.
\end{align}
For the $\pi^+\pi^-$ final state, the helicity angle $\theta$ is defined as the
angle between the 3-momentum of the $\pi^+$ in the rest frame of the $\pi\pi$
system and that of the $\pi\pi$ system in the rest frame of the initial
$\Upsilon(nS)$, where $\cos\theta\in [-1,1]$. The helicity angle for the
$\pi^0\pi^0$ final state is defined similarly; however, due to the indistinguishability of the two neutral pions,
we take $\cos\theta\in[0,1]$~\cite{CLEO2007}.
$t$ and $u$ can be expressed in terms of $s$ and $\theta$ according to
\begin{align}
t &= \frac{1}{2} \left[3s_0-s+\kappa(s)\cos\theta \right]\,,&
u &= \frac{1}{2} \left[3s_0-s-\kappa(s)\cos\theta \right]\,, \nn\\
\kappa(s) &\equiv \sigma_\pi \lambda^{1/2}\big(m_{\Upsilon(nS)}^2,m_{\Upsilon(mS)}^2,s\big) \,, &
\sigma_\pi &\equiv \sqrt{1-\frac{4m_\pi^2}{s}} \,,
\label{eq:tu}
\end{align}
where $\lambda(a,b,c)=a^2+b^2+c^2-2(ab+ac+bc)$.
We define $\vec{q}$  as the 3-momentum of the final vector meson in
the rest frame of the initial state with
\be
|\vec{q}|=\frac{1}{2m_{\Upsilon(nS)}}
\lambda^{1/2}\big(m_{\Upsilon(nS)}^2,m_{\Upsilon(mS)}^2,s\big) \,.
\ee

The results of the QCD multipole expansion together with the
soft-pion theorem can be reproduced by constructing a chiral
effective Lagrangian for the $\Upsilon(nS) \to \Upsilon(mS) \pi\pi$
transition. Since the spin of the heavy quarks decouples in the
heavy-quark limit, it is convenient to express the heavy quarkonia
in term of spin multiplets, and one has $J \equiv \vec{\Upsilon}
\cdot \boldsymbol{\sigma}+\eta_b$, where $\boldsymbol{\sigma}$
contains the Pauli matrices and $\vec\Upsilon$ and $\eta_b$
annihilate the $\Upsilon$ and $\eta_b$ states, respectively (see,
e.g., Ref.~\cite{Guo2011}). For the contact
$\Upsilon\Upsilon^{\prime}\pi\pi$ interaction, the effective
Lagrangian to leading order in the chiral as well as the heavy-quark
nonrelativistic expansion reads~\cite{Mannel}
\begin{equation}\label{LagrangianUpUppipi}
\L_{\Upsilon\Upsilon^{\prime}\pi\pi}
= \frac{c_1}{2}\bra J^\dagger J^\prime \ket \bra u_\mu u^\mu\ket
+\frac{c_2}{2}\bra J^\dagger J^\prime \ket \bra u_\mu u_\nu\ket v^\mu v^\nu
+\mathrm{h.c.} \,,
\end{equation}
where $v^\mu=(1,\vec{0})$ is the velocity of the heavy quark.\footnote{A further
chirally invariant term $\frac{c_0}{2}\bra J^\dagger J^\prime \ket \bra \chi_+ \ket+\mathrm{h.c.}$,
with $\chi_+  = u^\dagger   \chi u^\dagger + u \chi^\dagger u$,
$\chi=2 B \, {\rm diag}(m_u,m_d) + \ldots$,
includes a term $\propto B(m_u+m_d) \Upsilon^\dagger \Upsilon'+\mathrm{h.c.}$,
which will be eliminated upon diagonalization of the mass matrix for the $\Upsilon$ and $\Upsilon'$
states and therefore cannot contribute to the decay amplitude.}
The pions as Goldstone bosons of the spontaneous breaking of the approximate
chiral symmetry can be parametrized according to
\be
u_\mu = i \left( u^\dagger \partial_\mu u\, -\, u \partial_\mu u^\dagger\right) \,, \qquad
u^2 = e^{i {\Phi}/{ F_\pi}}\,, \qquad
\Phi =
\begin{pmatrix}
   \pi ^0  & \sqrt2{\pi^+ }  \\
   \sqrt2{\pi^- } & -\pi ^0  \\
\end{pmatrix} ,
\ee
where $F_\pi=92.2\MeV$ denotes the pion decay constant.

We need to define a $Z_b \Upsilon\pi$ interaction Lagrangian to calculate
the contribution of the virtual intermediate $Z_b$ states, $\Upsilon(nS)
\to Z_b\pi \to\Upsilon(mS) \pi\pi$.  The leading-order term
is proportional to the pion energy~\cite{Guo2011},
\be\label{LagrangianZbUppi}
\L_{Z_b\Upsilon\pi} = C_Z \Upsilon^i \bra {Z^i}^\dagger
u_\mu \ket v^\mu +\mathrm{h.c.} \,,
\ee
where
\begin{equation}
Z^i=
 \left( {\begin{array}{*{2}c}
   \frac{1}{\sqrt{2}}Z^{0i} & Z^{+i}   \\
   Z^{-i} & -\frac{1}{\sqrt{2}}Z^{0i}  \\
\end{array}} \right)\,.
\end{equation}
In the following, we will use $Z_{b1}$ and $Z_{b2}$ to refer to $Z_b(10610)$ and
$Z_b(10650)$, respectively, and use $C_{Z_{bi}\Upsilon(nS)\pi}$ to denote the
coupling constants for the $Z_{bi}\Upsilon(nS)\pi$ vertices.

We briefly comment on the mass dimensions of the LECs and coupling constants in
Eqs.~\eqref{LagrangianUpUppipi} and \eqref{LagrangianZbUppi}.
As the fields for the bottomonia and the $Z_b$ states are treated nonrelativistically,
in principle they should be normalized in a  nonrelativistic manner, leading to fields of mass dimension 3/2.
The difference to the usual relativistic normalization is a factor of $\sqrt{M}$,
with $M$ the mass of the heavy particle; since this difference is only a
constant, we choose to absorb it into the definition of
the coupling constants in the Lagrangians for simplicity, so that the heavy fields carry the
usual relativistic normalization instead.
Thus, the $c_i$ are dimensionless, while the $C_Z$ have mass dimension 1.

Note furthermore that, in order to preserve the analytic structure
of the amplitudes exactly, we keep fully relativistic propagators
for the $Z_b$ exchange graphs.

\begin{figure}
\centering
\includegraphics[width=0.7\linewidth]{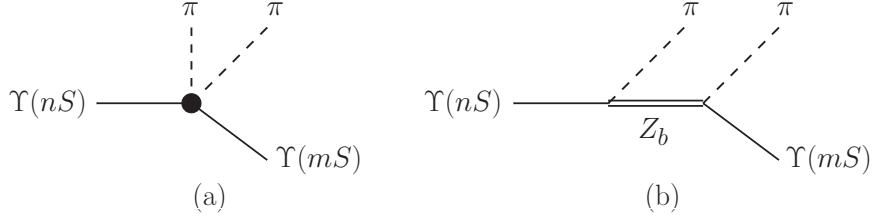}
\caption{Tree diagrams relevant to the decays $\Upsilon(nS) \to \Upsilon(mS) \pi \pi $:
(a) contact terms induced by the chiral Lagrangian; (b) $Z_b$ pole graphs.
The crossed pole diagram is not shown explicitly.}\label{fig.UpsnStoUpsmSpipiTree}
\end{figure}
The widths of the $Z_b$ states are of the order of $10\MeV$ and are
much smaller than the difference between the $Z_b$ masses and the
$\Upsilon(nS)\pi$ thresholds. Thus, they can be safely neglected in
the processes under investigation. Using the effective Lagrangians
in Eqs.~\eqref{LagrangianUpUppipi} and~\eqref{LagrangianZbUppi}, the
tree amplitude of the $\Upsilon(nS) \to \Upsilon(mS) \pi \pi$
processes as shown in Fig.~\ref{fig.UpsnStoUpsmSpipiTree} can be
written as
\begin{equation}
\label{eq.TreeAmplitude}
\M^\text{tree}(s,t,u)
= -\frac{4}{F_\pi^2}\epsilon_{\Upsilon(nS)}\cdot
\epsilon_{\Upsilon(mS)}\Bigg[ c_1 p_c\cdot p_d +c_2 p_c^0 p_d^0
+\sum_{i=1,2}\frac{C_{nm,i}}{2}\,  p_c^0
p_d^0\bigg(\frac{1}{t-m_{Z_{bi}}^2}+\frac{1}{u-m_{Z_{bi}}^2}\bigg)
\Bigg] ,
\end{equation}
where $\epsilon_{\Upsilon(nS)}$ and $\epsilon_{\Upsilon(mS)}$ are
polarization vectors, $p_c^0$ and $p_d^0$ denote the energies of
the pions in the lab frame, and $C_{nm,i}\equiv C_{Z_{bi}\Upsilon(nS)\pi} C_{Z_{bi}\Upsilon(mS)\pi}$
is the product of the coupling constants for the
exchange of the $Z_{bi}$. Here, we have neglected terms suppressed
by $p_c p_d/m_{Z_{bi}}^2$.

The partial-wave decomposition of $\M^\text{tree}$ can be easily performed by using
Eq.~\eqref{eq:tu} as well as the relation
\begin{equation}
p_c^0 p_d^0 =
\frac{1}{4} \left(s+\vec{q}^2\right)-
\frac{1}{4}\vec{q}^2 \sigma_\pi^2 \cos^2\theta \, .
\end{equation}
In view of the following treatment of pion-pion FSIs using
dispersive methods, it is useful to further decompose the partial
waves into contact terms derived from the chiral Lagrangian
Eq.~\eqref{LagrangianUpUppipi}, $M_l^\chi(s)$, and the projected
$Z_b$-exchange terms, $\hat M_l(s)$, in the form \be
\M^\text{tree}(s,\cos\theta) = \epsilon_{\Upsilon(nS)}\cdot
\epsilon_{\Upsilon(mS)}
\sum_{l=0}^{\infty}\left[M_l^\chi(s)+\hat{M}_l(s)\right]
P_l(\cos\theta)\,, \ee where $P_l(\cos\theta)$ are the standard
Legendre polynomials. Since parity conservation (or isospin
conservation combined with Bose symmetry) requires the pions to have
even relative angular momentum $l$, only even partial waves
contribute, and we only take into account the $S$- and $D$-wave
components in this study, neglecting the effects of yet higher
partial waves. Explicitly, the two parts of the  $S$-wave projection
of the tree amplitude read
\begin{align}\label{eq.M0chiral}
M_0^{\chi}(s)&= -\frac{2}{F_\pi^2}
\bigg\{c_1 \left(s-2m_\pi^2 \right)
+\frac{c_2}{2} \bigg[s+\vec{q}^2\Big(1  -\frac{\sigma_\pi^2}{3} \Big)\bigg]\bigg\} , \\
\hat{M}_0(s)&=-\frac{2}{F_\pi^2 \kappa(s)}\sum_{i=1,2}C_{nm,i} \Big\{
\left(s+ \vec{q}^2\right)Q_0(y_i)- \vec{q}^2
\sigma_\pi^2 \left[y_i^2 Q_0(y_i)-y_i\right] \Big\}\nonumber\\
&\equiv  \sum_{i=1,2}C_{nm,i}\, \bar{M}_{0i}(s)\,,\label{eq.M0hat}
\end{align}
where $y_i \equiv (3s_0-s-2m_{Z_{bi}}^2)/\kappa(s)$, and
$Q_0(y)$ is a Legendre function of the second kind,
\begin{equation}
Q_0(y)=\frac{1}{2}\int_{-1}^1 \frac{\diff z}{y-z}P_0(z)
 = \frac{1}{2}\log \frac{y+1}{y-1} \, .
\end{equation}
The $D$-wave projections are given by
\begin{align}\label{eq.M2chiral}
M_2^{\chi}(s)&=\frac{2}{3 F_\pi^2}c_2 \vec{q}^2 \sigma_\pi^2\,,\\
\hat{M}_2(s) &= -\frac{5}{F_\pi^2 \kappa(s)}\sum_{i=1,2}
C_{nm,i} \left(s+\vec{q}^2- \vec{q}^2 \sigma_\pi^2 y_i^2 \right)
\left[(3y_i^2-1)Q_0(y_i)-3y_i \right]\nonumber\\
&\equiv \sum_{i=1,2}C_{nm,i}\, \bar{M}_{2i}(s)\,.\label{eq.M2hat}
\end{align}

\subsection{Final-state interactions with dispersion relations}
\label{sect.dispersion}

The $\pi\pi$ system undergoes strong FSIs
in particular in the isospin-$0$ $S$-wave already at rather moderate energies
above threshold, which has to be included in a
theoretical calculation. A model-independent method to take FSIs into account is
given by dispersion theory. Based on unitarity and analyticity, it determines
the amplitudes up to certain subtraction constants, which can be obtained by
matching to the results of chiral effective theory.
For the processes $\Upsilon(nS) \to \Upsilon(mS) \pi\pi$ $(m < n\leq 3)$ studied
here, the invariant mass of the pion pair is well below the $K\bar K$ threshold.
Thus, it is not necessary to consider multichannel rescattering effects
explicitly.\footnote{We have checked that including the $K\bar K$ channel in a
two-channel Muskhelishvili--Omn\`es formalism (see Ref.~\cite{Daub} for an
application in the context of heavy-meson decays, as well as references therein)
would not lead to any significant change in our numerical results.}

We write the partial-wave expansion of the full
amplitude\footnote{In accordance with the tree-level amplitude, we
neglect all terms with other contractions of the polarization
vectors, which are suppressed in the heavy-quark nonrelativistic
expansion. } including the $\pi\pi$ FSI according to \be
\M^\text{full}(s,\cos\theta) =\epsilon_{\Upsilon(nS)}\cdot
\epsilon_{\Upsilon(mS)} \sum_{l=0}^{\infty}
\left[M_l(s)+\hat{M}_l(s)\right] P_l(\cos\theta)\,. \ee Here,
$M_l(s)$ contains the right-hand cut and accounts for $s$-channel
rescattering. On the other hand, $\hat{M}_l(s)$ represents
(partial-wave projected) left-hand cut contributions, be it due to
crossed-channel pole terms or rescattering effects. In the present
study, we approximate the left-hand cuts by $Z_b$ exchange only. The
functions $\hat{M}_l(s)$ are therefore given precisely by the
expressions in Eqs.~\eqref{eq.M0hat} and \eqref{eq.M2hat} already
quoted in the previous section. By construction, they are real and
free of discontinuities along the right-hand cut, such that in the
regime of elastic $\pi\pi$ rescattering, the partial-wave unitarity
conditions read
\begin{equation}\label{eq.unitarity}
\textrm{Im}\, M_l(s)= \left[M_l(s)+\hat{M}_l(s)\right]
\sin\delta_l^0(s) e^{-i\delta_l^0(s)}\,.
\end{equation}
Below the inelastic threshold (here the $K\bar{K}$ threshold), the
phases of the partial-wave amplitude $\delta_l^0$, of isospin $I=0$
and angular momentum $l$, coincide with the $\pi\pi$ elastic phase
shifts, as required by Watson's theorem~\cite{Watson1,Watson2}. To
solve Eq.~\eqref{eq.unitarity}, first we define the Omn\`es
function~\cite{Omnes},
\begin{equation}\label{Omnesrepresentation}
\Omega_l^I(s)=\exp
\bigg\{\frac{s}{\pi}\int^\infty_{4m_\pi^2}\frac{\diff x}{x}
\frac{\delta_l^I(x)}{x-s}\bigg\}\,,
\end{equation}
which obeys
$\Omega_l^I(s+i\epsilon)=e^{2i\delta_l^I}\Omega_l^I(s-i\epsilon)$.
Then the discontinuity of the function $m_l(s)\equiv
M_l(s)/\Omega_l^0(s)$ can be obtained with the help of Eq.~\eqref{eq.unitarity} as
\begin{equation}\label{eq.mldiscontinuity}
\frac{m_l(s+i\epsilon)-m_l(s-i\epsilon)}{2i}=
\frac{M_l(s+i\epsilon)e^{-i\delta_l^0}-M_l(s-i\epsilon)e^{i\delta_l^0}}{2i
|\Omega_l^0|}=\frac{\sin \delta_l^0 \hat{M}_l}{|\Omega_l^0|}\,.
\end{equation}
From the dispersion relation for the function $m_l(s)$, we then
obtain the solution of Eq.~\eqref{eq.unitarity}~\cite{Leutwyler96}
\be\label{OmnesSolution}
M_l(s)=\Omega_l^0(s)\bigg\{P_l^{n-1}(s)+\frac{s^n}{\pi}\int_{4m_\pi^2}^\infty
\frac{\diff x}{x^n} \frac{\hat
M_l(x)\sin\delta_l^0(x)}{|\Omega_l^0(x)|(x-s)}\bigg\} \,, \ee where
the polynomial $P_l^{n-1}(s)$ is a subtraction function. In the
absence of the inhomogeneous terms $\hat{M}_l(s)$ in the unitarity
condition, i.e.\ without left-hand cuts, we would have found a
standard Omn\`es solution for a form factor $P_l^{n-1}(s)\,
\Omega_l^I(s)$, which is valid in the case where
\begin{figure}
\includegraphics[width=0.7\linewidth]{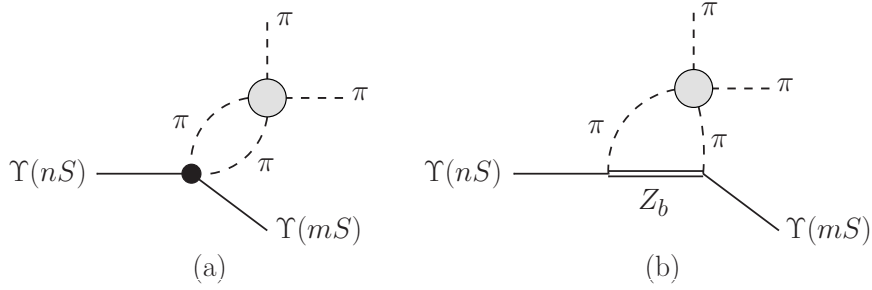}
\caption{Pion-pion final-state interactions (a) with the two pions
originating from a point source, denoted by the black dot, and (b)
with pions produced by a $Z_b$ pole term.  The gray blob denotes
pion-pion rescattering.} \label{fig:FSI}
\end{figure}
the production of the two pions can be thought to originate from a
point source; see Fig~\ref{fig:FSI}\,(a). The modified solution in
Eq.~\eqref{OmnesSolution} contains a dispersion integral over the
inhomogeneities $\hat{M}_l(s)$, which represents the rescattering
including the production from a pole term, see
Fig~\ref{fig:FSI}\,(b), and provides the crossed-channel $Z_b$
exchange graph with the correct phase in accordance with Watson's
theorem. Very similar methods to include resonance exchange as an
approximation to left-hand-cut structures have been applied recently
to processes such as
$\gamma\gamma\to\pi\pi$~\cite{Moussallam-gamma},
$\eta\to\pi\pi\gamma$~\cite{KubisPlenter}, or $B \to \pi\pi l
\nu_l$~\cite{Kang}.

In order to determine the necessary number of subtractions in Eq.~\eqref{OmnesSolution},
we need to make sure that the dispersive integral over the inhomogeneities converges,
and hence have to investigate the high-energy behavior of the integrand.
We first remark that
for a phase shift $\delta_l^I(s)$ reaching $k\,\pi$ at high
energies, the corresponding Omn\`es function falls off asymptotically as $s^{-k}$.
 Assuming that both the $S$-wave and $D$-wave $\pi\pi$ scattering phase shifts,
$\delta_{0,2}^0(s)$, approach $\pi$ for high energies, we have
$\Omega_{0,2}^0(s) \sim 1/s$ for large $s$. Second, we have checked
that in an intermediate energy range of $1\GeV^2 \lesssim s \ll
m_{\Upsilon}^2$, both inhomogeneities grow at most linearly in $s$.
We conclude that in the dispersive representations for $M_0(s)$ and
$M_2(s)$, three subtractions are sufficient to render the dispersive
integrals convergent.

At low energies, i.e.\ close to or even below threshold, $M_0(s)$ and $M_2(s)$ can be matched to the chiral representation.
We perform the matching in the limit of $\pi\pi$ rescattering being switched off, i.e.\ $\Omega_l^0(s)\equiv 1$,
so that the subtraction functions can be identified exactly with the expressions given in
Eqs.~\eqref{eq.M0chiral} and~\eqref{eq.M2chiral}.
As both $M_0^\chi(s)$ and $M_2^\chi(s)$ grow no faster than $\sim s^2$,
the degree of the subtraction polynomial covers these terms.
Therefore, the integral equations take the form
{\allowdisplaybreaks
\begin{align}
M_0(s)&=\Omega_0^0(s)\bigg\{-\frac{2}{F_\pi^2}
\bigg[ c_1 \left(s-2m_\pi^2 \right)
+\frac{c_2}{2} \bigg(s+\vec{q}^2\Big(1  -\frac{\sigma_\pi^2}{3} \Big)\bigg)\bigg] \nn\\ & \qquad \qquad
+\sum_{i=1,2}C_{nm,i} \frac{s^3}{\pi}\int_{4m_\pi^2}^{\infty} \frac{\diff x}{x^3}
\frac{\bar{M}_{0i}(x)\sin\delta_0^0(x)}{|\Omega_0^0(x)|(x-s)}
\bigg\}\,, \nonumber\\
M_2(s)&=\Omega_2^0(s)\bigg\{\frac{2}{3 F_\pi^2}c_2
\vec{q}^2\sigma_\pi^2
+\sum_{i=1,2}C_{nm,i}\, \frac{s^3}{\pi}\int_{4m_\pi^2}^{\infty} \frac{\diff x}{x^3}
\frac{\bar{M}_{2i}(x)\sin\delta_2^0(x)}{|\Omega_2^0(x)|(x-s)}
\bigg\} \,. \label{eq.M02}
\end{align}
}%
A subtlety in this prescription concerns the kinematically singular
parts of the subtraction functions $\propto 1/s$ that derive from
the similarly singular inhomogeneities: the subtractions functions
in Eq.~\eqref{eq.M02} are not actually subtraction
\textit{polynomials}. These are an artifact of the partial-wave
decomposition: the complete (polynomial) chiral amplitude as
contained in Eq.~\eqref{eq.TreeAmplitude} is obviously nonsingular,
and due to $\Omega_0^0(0)=\Omega_2^0(0)=1$, this cancellation in the
combination of partial waves is preserved in the dispersive
representation. We show how to argue for the
representation~\eqref{eq.M02} more rigorously in
Appendix~\ref{Appendix.A}.

It is then straightforward to calculate the $\pi\pi$ invariant mass spectrum and
helicity angular distribution for $\Upsilon(nS) \to \Upsilon(mS)
\pi^+\pi^-$ using
\begin{equation}
\frac{\diff\Gamma}{\diff \sqrt{s} \,\diff\cos\theta} =
\frac{\sqrt{s}\,\sigma_\pi |\vec{q}|}{128\pi^3 m_{\Upsilon(nS)}^2}
\left|M_0+\hat{M}_0+(M_2+\hat{M}_2)
P_2(\cos\theta)\right|^2\,,\label{eq.pipimassdistribution}
\end{equation}
where we have made use of
$\sum_{\lambda,\lambda'} \big|\epsilon_{\Upsilon(nS)}^{(\lambda)}\cdot
\epsilon_{\Upsilon(mS)}^{(\lambda')}\big|^2 \approx 3$, which is
an approximation accurate to a few per mil.
For the neutral-pion process $\Upsilon(nS) \to \Upsilon(mS)
\pi^0\pi^0$, Eq.~\eqref{eq.pipimassdistribution} needs to be multiplied by
$1/2$ in order to
account for the indistinguishable neutral pions in the final state.

\section{Phenomenological discussion} \label{pheno}

We first discuss the $\pi\pi$ phase shifts used in the calculation
of the Omn\`es functions and the dispersion integrals. As we
describe the $S$-wave in a single-channel approximation, i.e.\
without taking inelasticities due to $K\bar K$ intermediate states
into account explicitly, we employ the phase of the nonstrange pion
scalar form factor (as determined in Ref.~\cite{Hoferichter:2012wf}
from the solution of the coupled-channel Muskhelishvili--Omn\`es
problem) instead of $\delta_0^0$, which yields a good description at
least below the onset of the $K\bar K$ threshold. For the $D$-wave,
we use the parametrization for $\delta_2^0$ given by the
Madrid--Krak\'ow collaboration~\cite{Pelaez}. Both phases are guided
smoothly to the assumed asymptotic values $\delta_0^0(s),$
$\delta_2^0(s) \to \pi$ for $s\to\infty$. In practice, the
dispersion integrals over the inhomogeneities in Eq.~\eqref{eq.M02}
are cut off at $s = (3\GeV)^2$; above that point, the phases are so
close to $\pi$ already that the contributions to the dispersive
integrals in Eq.~\eqref{eq.M02} can be neglected.

All the LECs in the chiral Lagrangian Eq.~\eqref{LagrangianUpUppipi}
are unknown, and will be fitted to the experimental data for the
$\Upsilon(nS)\to\Upsilon(mS)\pi\pi$ transitions. These LECs are
different for processes with different values of $n$ and $m$, since
there is no symmetry connecting different radial excitations of the
bottomonium states. The experimental data that we will use include
the $\pi\pi$ invariant mass distributions and the helicity angular
distributions  for the $\Upsilon(nS) \to \Upsilon(mS) \pi\pi$
$(m<n\leq3)$ processes measured by the CLEO Collaboration in
Ref.~\cite{CLEO2007}. For the transitions from $\Upsilon(3S)$ to
$\Upsilon(1S)$ and from $\Upsilon(2S)$ to $\Upsilon(1S)$, we
simultaneously fit to the data of the $\pi^0\pi^0$ and the
$\pi^+\pi^-$ final state. For the transition from $\Upsilon(3S)$ to
$\Upsilon(2S)$, we only fit the data of the $\pi^0\pi^0$ final state
due to the limited statistics of the $\Upsilon(3S) \to \Upsilon(2S)
\pi^+\pi^-$ process (the event number is almost 1 order of magnitude
smaller than the one for the $\pi^0\pi^0$ channel).

In principle, the
$Z_{bi}\Upsilon(nS)\pi$ coupling strengths can be extracted from measuring the
partial widths of both $Z_b$ states into $\Upsilon(nS)\pi$ $(n\leq3)$ using
\begin{equation}\label{eq.CZ}
|C_Z|=\Bigg\{\frac{4\pi F_\pi^2 m_{Z_b}^2 \Gamma_{Z_b \to
\Upsilon\pi}}{|\vec{p}_f|\big(m_\pi^2+\vec{p}_f ^2\big)}
\Bigg\}^{\frac{1}{2}} ,
\end{equation}
where $|\vec{p}_f|\equiv
\lambda^{1/2}\big(m_{Z_b}^2,m_\Upsilon^2,m_\pi^2\big)/(2m_{Z_b})$, and
$\Gamma_{Z_b \to \Upsilon\pi}$ is the partial width for the $Z_b\to\Upsilon\pi$
decay. Thus, the coupling strengths can be obtained if the partial widths are
known. In fact, there are preliminary results for the branching fractions of the
decays of both $Z_b$ states into $\Upsilon(nS)\pi$
$(n\leq3)$~\cite{Belle2012:2}, where the $Z_b$ line shapes were described using
Breit--Wigner forms.
All branching fractions are found to be of the order of a few per cent.
If we naively calculated the partial widths by multiplying these branching fractions by the measured width of the $Z_b$ states,
we would obtain\footnote{The branching fractions for $Z_b(10650)$ decays in
Table~V of Ref.~\cite{Belle2012:2} are divided by 1.33, as mentioned at the end of the
experimental paper, to account for the decay mode $Z_b(10650)\to B\bar B^*$.}
\begin{align}\label{eq.CZvalue}
|C_{Z_{b1} \Upsilon(1S)\pi}^\text{naive}|&=0.024 \pm 0.003, &
|C_{Z_{b1} \Upsilon(2S)\pi}^\text{naive}|&=0.23 \pm 0.03, &
|C_{Z_{b1} \Upsilon(3S)\pi}^\text{naive}|&=0.60 \pm 0.08, \nonumber\\
|C_{Z_{b2} \Upsilon(1S)\pi}^\text{naive}|&=0.013 \pm 0.002, &
|C_{Z_{b2} \Upsilon(2S)\pi}^\text{naive}|&=0.11 \pm 0.01, &
|C_{Z_{b2} \Upsilon(3S)\pi}^\text{naive}|&=0.28 \pm 0.03
\end{align}
(all in units of GeV),
and the products of couplings relevant for the process
$\Upsilon(3S)\to\Upsilon(1S)\pi\pi$ are
\begin{equation}\label{eq.CZCZvalue}
|C_{31,1}^\text{naive}|=(0.014 \pm 0.004)\GeV^2,\qquad |C_{31,2}^\text{naive}|=
(0.004 \pm 0.001)\GeV^2\,.
\end{equation}
Here all the extractions are labeled by a superscript ``naive'' because this is
not the appropriate way of extracting the coupling strengths in this case: the $Z_b$
structures are very close to the $B^{(*)}\bar B^*$ thresholds, and thus a
Flatt\'e parametrization should be used, which will lead to much larger partial
widths into $\Upsilon\pi$ (and $h_b\pi$), and thus the relevant coupling
strengths. As discussed in Appendix~\ref{Appendix.B}, the sum of the partial
widths of the $Z_b(10610)$ other than that for the $B\bar B^*$ channel should be
larger than the nominal width, which is about $20\MeV$. This would require at
least some of the couplings to the $(b\bar b)\pi$ channels to
be significantly larger than the values indicated by naive calculation using
branching fractions. Taking the $Z_b(10610)$ as an example, summing over all the
$\Upsilon(nS)\pi\,(n=1,2,3)$ and $h_b(mP)\pi\,(m=1,2)$ branching fractions in
Ref.~\cite{Belle2012:2} gives about 14\% or $3\MeV$ in terms of partial widths. We
therefore expect $|C_{Z_{b1}\Upsilon(nS)\pi}|^2$ to be roughly
one order of magnitude larger than those from Eq.~\eqref{eq.CZvalue},\footnote{
The extraction of these coupling constants using a Flatt\'e-like parametrization
requires a detailed analysis of the data for all the mentioned $Z_b$ decay
channels, and is beyond the scope of this paper.
We notice that such a procedure was recently proposed in
Ref.~\cite{Hanhart:2015cua}.} and thus
\begin{equation}\label{eq.CZCZvalue1}
|C_{31,1}|= \mathcal{O}(0.1\,\GeV^2)\, .
\end{equation}

\begin{figure}
 \begin{center}
  \includegraphics[width=0.9\textwidth]{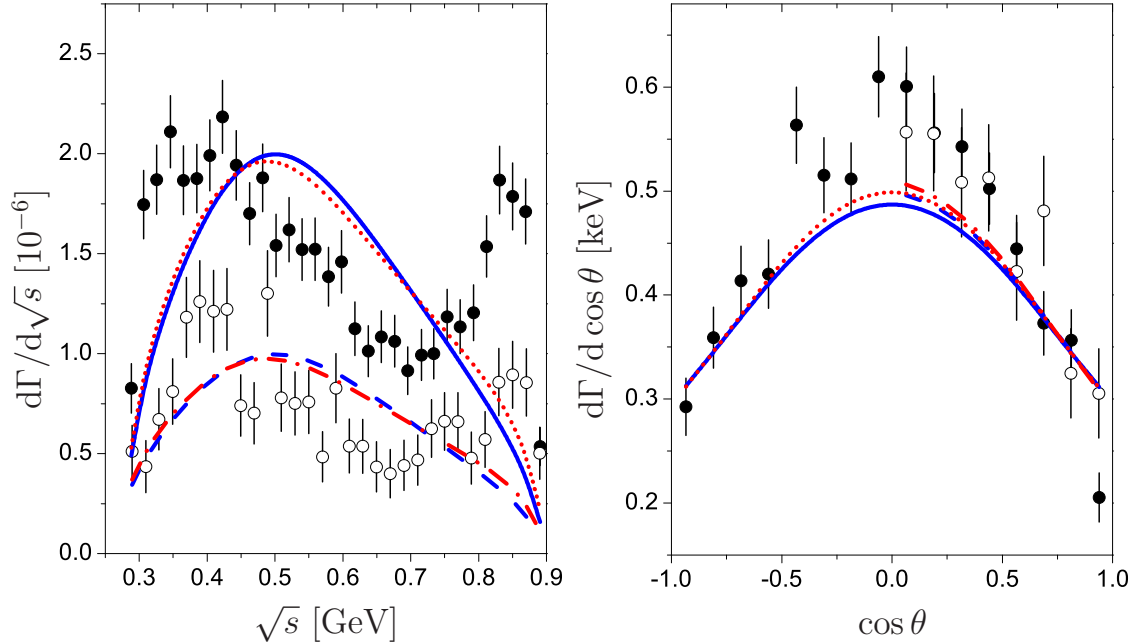}
 \end{center}
 \caption{Simultaneous fit to the $\pi\pi$ invariant mass distributions and the
 helicity angle distributions in $\Upsilon(3S) \to \Upsilon(1S)\pi\pi$. The
 solid (dashed) and dotted (dot-dashed) curves show the best fit results without considering the $Z_b$ and using
 central values of the $Z_{bi}\Upsilon(nS)\pi$ couplings,  given in
 Eq.~\eqref{eq.CZvalue}, extracted from the $Z_b$ branching fractions.
 Solid and open circles refer to data points for the charged- and neutral-pion final states, respectively.
 }
 \label{fig:simfit}
\end{figure}
Because $\Upsilon(3S)\to\Upsilon(1S)\pi\pi$ is of particular interest for
its unusual shape of the dipion invariant mass distribution, we will focus on
this decay mode first. We try to fit to the dipion invariant mass
distribution and the helicity angular distribution simultaneously without
including any of the $Z_b$ states.
The results of the best fit are shown as the solid (dashed) curves for the
$\pi^+\pi^-$ ($\pi^0\pi^0$) mode in Fig.~\ref{fig:simfit}. It is obvious that
the double-bump behavior of the invariant mass spectrum is not reproduced,
although the angular distribution is described well. This calls for a new
mechanism in addition to the $\pi\pi$ FSIs. We then include both
$Z_b$ states. Since the coupling constants for the $Z_b\Upsilon\pi$ vertices
extracted using the Flatt\'e form are not available, we try to fix them to the
central values in Eq.~\eqref{eq.CZCZvalue}. The
results are shown as the dotted (dot-dashed) curves in the same figure.
Obviously, the best fits in both cases are very similar to each
other.

\begin{figure}
\centering
\includegraphics[width=\linewidth]{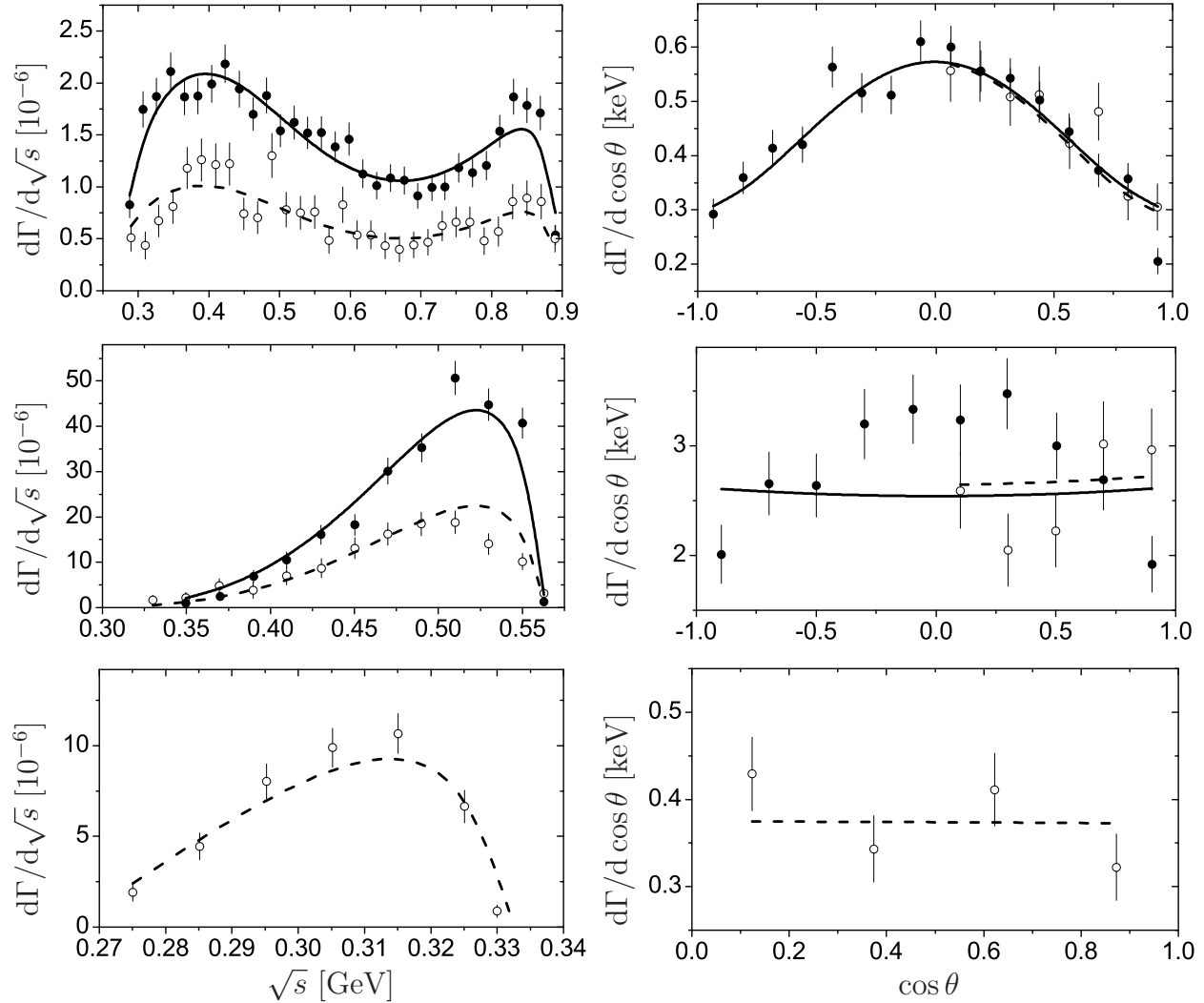}
\caption{Fit results for the decays $\Upsilon(3S) \to \Upsilon(1S)
\pi\pi $, $\Upsilon(2S) \to \Upsilon(1S) \pi\pi $, and $\Upsilon(3S)
\to \Upsilon(2S) \pi^0\pi^0 $ (from top to bottom). The left panels
display the $\pi\pi$ invariant mass spectra, while the right panels
show the $\cos\theta$ distributions. The solid and open circles
denote the charged and neutral decay mode data, respectively; full
and dashed lines show the theoretical fit results for charged- and
neutral-pion final states.} \label{fig.MassAngularDistributions}
\end{figure}

It is interesting to see what happens if we treat the couplings of
the $Z_b$ states to the $\Upsilon\pi$ as free parameters as well.
However, the mass difference between the two $Z_b$ states, about
only $40\MeV$, is much smaller than the gap between their masses and
the $\Upsilon(nS)\pi$ $(n=1,2,3)$ thresholds; they have the same
quantum numbers and thus the same coupling structure as dictated by
Eq.~\eqref{LagrangianZbUppi}.  It is therefore very difficult to
distinguish their effects from each other in the processes under
investigation. In practice, this means that the couplings for the
$Z_{b}(10610)$ and $Z_b(10650)$ are strongly correlated in the fit
and it is impossible to obtain a sensible uncertainty for them.
Therefore, we use only one $Z_b$ state, by setting $C_{nm,2}=0$, and
take its mass to be that of the $Z_b(10610)$. With three free
parameters $c_1$, $c_2$, and $C_{31,1}$, we are able to achieve a
very good agreement with the data for both the invariant mass and
helicity angular distribution, as can be seen from the upper panel
of Fig.~\ref{fig.MassAngularDistributions}.
In addition, the data for the processes
$\Upsilon(2S)\to\Upsilon(1S)\pi\pi$ and
$\Upsilon(3S)\to\Upsilon(2S)\pi\pi$ are also fitted, shown as the
middle and lower panels in Fig.~\ref{fig.MassAngularDistributions},
respectively. It is not surprising that the invariant mass
distributions for both of these two processes are described well, as
their phase spaces are not large enough to allow for nontrivial
structures comparable to the one for $\Upsilon(3S)\to\Upsilon(1S)\pi\pi$.
Still, the agreement with the data for the angular distribution for
$\Upsilon(2S)\to\Upsilon(1S)\pi\pi$ is not as good. This is
mainly because of the discrepancy between the data for the modes
with charged and neutral pions. This discrepancy was
attributed to different efficiencies for reconstruction and
resolutions, as well as the folding of the neutral angle in the
experimental paper~\cite{CLEO2007}, which are not available and thus
not considered in our fit. The resulting values of the parameters as
well as the $\chi^2$ per degree of freedom are shown in
Table~\ref{tablepar1}.
\begin{table}
\caption{\label{tablepar1} The parameter results from the fits of
the $\Upsilon(nS) \to \Upsilon(mS) \pi\pi $ processes. For
the transitions from $\Upsilon(3S)$ to $\Upsilon(1S)$ and from
$\Upsilon(2S)$ to $\Upsilon(1S)$, we simultaneously fit the data of
the $\pi^0\pi^0$ final state and the $\pi^+\pi^-$ final state. For
the transition from $\Upsilon(3S)$ to $\Upsilon(2S)$, we only fit
the data of the $\pi^0\pi^0$ final state, due to the limited
statistics of the  $\Upsilon(3S) \to \Upsilon(2S)
\pi^+\pi^-$ process.}
\renewcommand{\arraystretch}{1.2}
\begin{center}
\begin{tabular}{lccc}
\toprule
         & $~\Upsilon(3S) \to \Upsilon(1S) \pi\pi ~$
         & $~\Upsilon(2S) \to \Upsilon(1S) \pi\pi ~$
         & $~\Upsilon(3S) \to \Upsilon(2S) \pi^0\pi^0 ~$\\
\hline
$c_1$   &   $ -0.025\pm 0.001$           & $\phantom{-}0.09\pm 0.05 $ & $ -0.6\pm 0.1$ \\
$c_2$   &   $ \phantom{-}0.026\pm 0.001$ & $\phantom{-}0.04\pm 0.08 $ & $ \phantom{-}0.2\pm 0.3$  \\
$ C_{nm,1}~[\text{GeV}^2]$ & $\phantom{-}0.145\pm 0.006$ &$\phantom{-}1.3\pm 1.4$&$ \phantom{-}3.7\pm 2.6$ \\
\hline
 $\frac{\chi^2}{\rm d.o.f}$ &  $\frac{108.18}{87-3}=1.29$   &
$\frac{101.68}{40-3}=2.75$&  $\frac{12.18}{11-3}=1.52$  \\
\botrule
\end{tabular}
\end{center}
\renewcommand{\arraystretch}{1.0}
\end{table}
Note that the fitting results are invariant under a sign change of
all parameters simultaneously, as can be seen from
Eq.~\eqref{eq.pipimassdistribution}. The resulting values of the
LECs $c_i$ are very different for different transitions. These
parameters are determined by short-distance physics, that is, the
structure of the involved $\Upsilon(nS)$ states. Thus, such a
difference may be explained by the node structures of different
radial bottomonium excitations~\cite{TMYan1980,Kuang1981}.  We also
notice that the node structure affects the coupling constants that
are determined by the internal bottomonium structure but do not have
an impact on the dipion invariant mass distribution.

We observe that the product of the $Z_b$ couplings to
$\Upsilon(3S)\pi$ and $\Upsilon(1S)\pi$, $C_{31,1}$, is well
constrained, while the values of $C_{21,1}$ and $C_{32,1}$ are
consistent with zero (within 1.5 standard deviations for the
latter). The  value of $C_{31,1}$ extracted in this way is 1 order
of magnitude larger than the naive value given in
Eq.~\eqref{eq.CZCZvalue}; however, it is of the same order as the
expectation in Eq.~\eqref{eq.CZCZvalue1}. Notice that we have
switched off the higher $Z_b$ in the fit, and thus the extracted
coupling constant should be understood as containing effects from
both $Z_b$ states.

\begin{figure}
\centering
\includegraphics[width=0.8\linewidth]{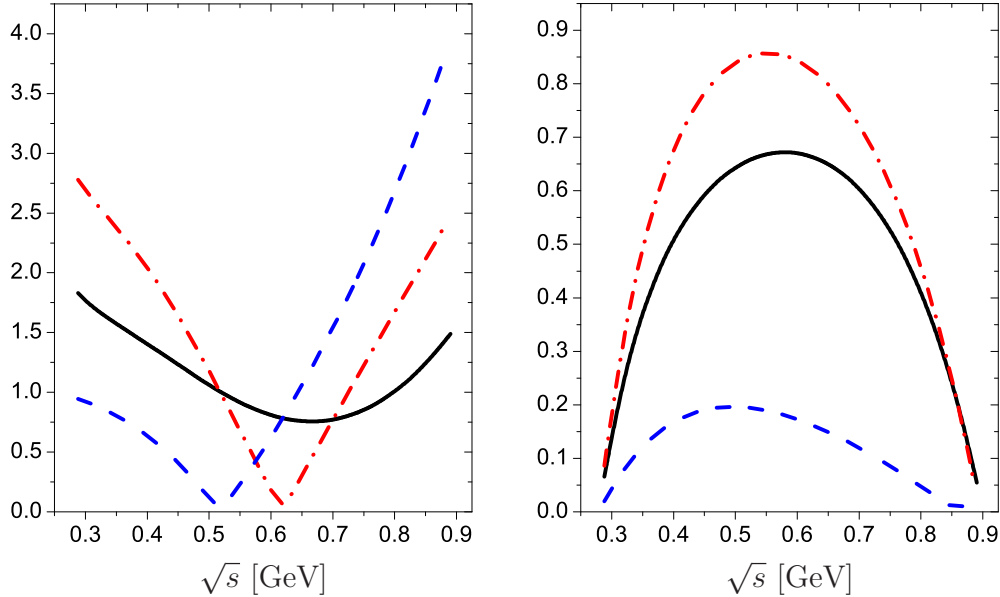}
\caption{Moduli of the $S$- (left) and $D$-wave (right)amplitudes in the
$\Upsilon(3S) \to \Upsilon(1S) \pi^+\pi^-$ process. The black solid lines
represent our best fit results, while the red dot-dashed and blue dashed lines correspond to the contributions from
the $c_i$ terms and the $Z_b(10610)$, respectively.}
\label{fig.SwaveDwaveAmplitudes}
\end{figure}

Since the value of $C_{31,1}$ is well constrained, it is instructive
to analyze different partial-wave components of the decay amplitude
for $\Upsilon(3S)\to\Upsilon(1S)\pi\pi$. In
Fig.~\ref{fig.SwaveDwaveAmplitudes}, we plot the moduli of the
$S$-wave and $D$-wave amplitudes from the $c_i$ terms and the
$Z_b(10610)$ state for this process. Notice that, while the $c_1$
term is a pure $S$-wave, the $c_2$ term contributes to both $S$- and
$D$-waves, and the $Z_b$-exchange in principle affects all partial
waves. One observes that the $D$-wave contribution from the
$Z_b$-exchange is much smaller than that from the $c_2$ term. This
means that the curved behavior of the observed angular distribution
is mainly due to the $c_2$ term. It should be mentioned that this
observation is different from the one in Ref.~\cite{Guo2005}, where
the intermediate tetraquark state, analogous to the $Z_b$ here, is
found to be dominant in the angular distribution. The reason is that
in Ref.~\cite{Guo2005} the mass of the tetraquark is fitted to
$10.08\GeV$, located between the masses of the $\Upsilon(3S)$ and
$\Upsilon(1S)$ states. If we fix the $Z_b$ mass to such a low value,
we indeed find that the ratio of the $D$- to $S$-wave components of
the pure $Z_b$-exchange mechanism significantly increases. For the
$S$-wave amplitudes, the contribution from the $c_i$ terms and that
from the $Z_b$-exchange are of the same order, and both of them have
a zero in the energy region of interest, responsible for the dip in
the invariant mass distribution.

\section{Conclusions}
\label{conclu}

We have used dispersion theory to study FSIs in the decays
$\Upsilon(nS) \to \Upsilon(mS) \pi\pi$, $(m < n \leq 3)$. In
particular,  we have analyzed the role of the $Z_b(10610)$ and
$Z_b(10650)$ states in these transitions. Pion-pion FSIs have been
considered in a model-independent way, and the leading chiral
amplitude acts as the subtraction function in the modified Omn\`{e}s
solution. Through fitting the data of the $\pi\pi$ mass spectra and
the angular $\cos\theta$ distributions, the couplings of the
$\Upsilon\Upsilon^{\prime}\pi\pi$ vertex as well as the product of
couplings of the $Z_b\Upsilon\pi$ vertex and the $Z_b
\Upsilon^\prime\pi$ vertex are determined. We find that the $Z_b$
effects in $\Upsilon(2S) \to \Upsilon(1S) \pi \pi $ and
$\Upsilon(3S) \to \Upsilon(2S) \pi^0 \pi^0 $ are very small, while
they play a significant role in the $\Upsilon(3S) \to \Upsilon(1S)
\pi \pi $ decay, which has a double-peak $\pi\pi$ mass spectrum. The
product of couplings $C_{Z_{b1} \Upsilon(3S)\pi} C_{Z_{b1}
\Upsilon(1S)\pi}$ obtained here is much larger than the one
extracted naively from the branching fractions of the
Breit--Wigner-parametrized $Z_b(10610)$ decays to $\Upsilon(nS)\pi^+
(n=1, 3)$ in Ref.~\cite{Belle2012:2}. It is, however, consistent
with a rough estimate based on a Flatt\'e parametrization for the
$Z_b(10610)$, which is in fact more appropriate for near-threshold
states. This analysis calls for a detailed study for the partial
widths of $Z_b(10610,10650)\to\Upsilon(1S,3S)\pi$ by analyzing the
data for $\Upsilon(5S)\to\Upsilon(1S,3S)\pi\pi$, together with other
processes where the $Z_b$ structures were observed, using, e.g., the
formalism presented in Ref.~\cite{Hanhart:2015cua}.

Therefore, our results show the necessity to analyze the dipion
decays of the $\Upsilon(nS)\,(n=3,4,5)$  states simultaneously,
taking into account all the effects from $\pi\pi$ strong FSIs, the
$Z_b$ states, and intermediate bottom mesons. The latter were
neglected here because the $\Upsilon(3S)$ is well below the $B\bar
B$ threshold and the left-hand-cut contribution due to the
$Z_b(10610)$, located near the $B^{(*)}\bar B^{(*)}$ thresholds,
could mimic the effects of the intermediate bottom mesons. Such a
combined study, taking pion-pion final-state interactions into
account consistently in the formalism laid out in this article,
while allowing for more general intermediate states as left-hand-cut
structures, should be pursued in the future. It would be most
valuable to finally understand the peculiar behavior of the
$\Upsilon(3S)\to\Upsilon(1S)\pi\pi$ decays on the one hand and to
learn more about the $Z_b$ structures on the other.

\section*{Acknowledgments}
We are grateful to Ling-Yun Dai, Christoph Hanhart, Xian-Wei Kang,
Roman Mizuk, De-Liang Yao, and  Han-Qing Zheng for helpful
discussions. This work is supported in part by NSFC and DFG through
funds provided to the Sino--German CRC110 ``Symmetries and the
Emergence of Structure in QCD'' (NSFC Grant No.~11261130311).
F.~K.~G. is also supported by NSFC (Grant No.~11165005) and by the
Thousand Talents Plan for Young Professionals. The work of U.~G.~M
was supported in part by the Chinese Academy of Sciences President's
International Fellowship Initiative (Grant No.\ 2015VMA076).

\appendix

\section{Singular inhomogeneities} \label{Appendix.A}

The contribution $\propto c_2$ in the tree amplitude in
Eq.~\eqref{eq.TreeAmplitude} has the property of yielding $S$- and $D$-wave projections
that diverge at $s=0$, while the combined expression is of course a polynomial in the
Mandelstam variables $s$, $t$, and $u$:
it can be written as (without changing the
essence of the issue, we leave out all polarization vectors and
overall prefactors such as coupling constants)
\begin{align}
s&+\vec{q}^2-\vec{q}^2 \Big(1-\frac{4m_\pi^2}{s}\Big)\cos^2\theta
= s+\vec{q}^2-\frac{1}{4m_{\Upsilon(nS)}^2}(t-u)^2\nonumber\\
&= s+\vec{q}^2 +\frac{s^2}{4m_{\Upsilon(nS)}^2} -
\frac{1}{2m_{\Upsilon(nS)}^2}\bigg\{
\Big[t^2-3s_0t+\frac{9}{4}s_0^2\Big] +
\Big[u^2-3s_0u+\frac{9}{4}s_0^2\Big] \bigg\}\,.\label{eq.c3}
\end{align}
In the main text, we have claimed that these singular partial-wave
projections can be included in a subtraction function of the Omn\`es
representation, although these clearly do not constitute a
subtraction \textit{polynomial}.  In this Appendix, we show how this
can be justified.

The two terms in the curly brackets of Eq.~\eqref{eq.c3} can be interpreted as (polynomial)
$S$-wave amplitudes in the $t$- and $u$-channels of the decay.
The projection of these
onto $s$-channel partial waves yields additional contributions $\delta  \hat
M_l(s)$, $l=0,2$, to the hat functions, on top of the terms
stemming from projected $Z_b$ pole terms. These additional contributions can be calculated easily:
\begin{align}
\delta\hat M_0(s) &\propto -\frac{1}{4m_{\Upsilon(nS)}^2}\int_{-1}^1 \diff\cos\theta
 \bigg\{
\Big[t^2-3s_0t+\frac{9}{4}s_0^2\Big] +
\Big[u^2-3s_0u+\frac{9}{4}s_0^2\Big] \bigg\} =
- \frac{\kappa^2(s)+3s^2}{12m_{\Upsilon(nS)}^2}  \,, \nonumber\\
\delta\hat M_2(s)&\propto  -\frac{5}{4m_{\Upsilon(nS)}^2}\int_{-1}^1 \diff\cos\theta P_2(\cos\theta)
\bigg\{
\Big[t^2-3s_0t+\frac{9}{4}s_0^2\Big] +
\Big[u^2-3s_0u+\frac{9}{4}s_0^2\Big] \bigg\}
= -\frac{\kappa^2(s)}{6m_{\Upsilon(nS)}^2} .
\end{align}
We note that, with $s \ll m_{\Upsilon(nS)}^2$, the term $\propto s^2$ can be neglected,
and we can use the approximation
\begin{equation}
\kappa^2(s) \approx
(m_{\Upsilon(nS)}+m_{\Upsilon(mS)})^2\Big[(m_{\Upsilon(nS)}-m_{\Upsilon(mS)})^2-s\Big]
\bigg(1-\frac{4m_\pi^2}{s} \bigg) \,,
\end{equation}
such that the
$\delta\hat M_l(s)$ only grow linearly with $s$ for large (but not
too large to be comparable with the $\Upsilon$ masses) energies.

With a polynomial inhomogeneity, the dispersive integral can be
performed analytically, using dispersive representations of the
inverse of the Omn\`es function (see, e.g.,~\cite{Hoferichter}); we
define
\begin{equation}
I_n(s) = - \frac{1}{\pi}\int_{4m_\pi^2}^\infty \frac{\diff x}{x^n}
\frac{\sin\delta(x)}{|\Omega(x)| (x-s)}\,
\end{equation}
and find
\begin{align}
\label{eq:invOmnes} \Omega^{-1}(s) &= 1 -
s\,\dot{\Omega}(0) + s^2 I_2(s) = 1 - s\,\dot{\Omega}(0) -
\frac{s^2}{2}\big[\ddot{\Omega}(0)-2\dot{\Omega}^2(0)\big]  + s^3
I_3(s)  \nn\\
&= 1 - s\,\dot{\Omega}(0) -
\frac{s^2}{2}\big[\ddot{\Omega}(0)-2\dot{\Omega}^2(0)\big]
-\frac{s^3}{6}\big[\dddot{\Omega}(0) -
6\ddot{\Omega}(0)\dot{\Omega}(0) + 6 \dot{\Omega}^3(0) \big] + s^4
I_4(s)
\end{align}
(assuming $\Omega(s) \sim 1/s$ for large
$s$), which can be solved for the $I_n(s)$. The full contribution of
the additional inhomogeneities to the partial-wave amplitudes is then given by
\begin{equation}
\delta \hat M(s) + \Omega(s) \frac{s^3}{\pi}
\int_{4m_\pi^2}^\infty \frac{\diff x}{x^3} \frac{\delta \hat
{M}(x)\sin\delta(x)}{|\Omega(x)| (x-s)} \,.
\end{equation}
If we write
$\delta\hat{M}(s) = m_1 s + m_0 + m_{-1}/s$, the terms involving
$\Omega^{-1}(s)$ in the solutions of Eq.~\eqref{eq:invOmnes} for
$I_n(s)$ exactly cancel $\delta\hat{M}(s)$. One ends up with a
partial-wave contribution
\begin{align}
\Omega(s) & \bigg\{  m_1s\big( 1 - s\,\dot{\Omega}(0) \big) + m_0
\Big(1 - s\,\dot{\Omega}(0) -
\frac{s^2}{2}\big[\ddot{\Omega}(0)-2\dot{\Omega}^2(0)\big] \Big)
\nn\\
& + \frac{m_{-1}}{s} \bigg(1 - s\,\dot{\Omega}(0) -
\frac{s^2}{2}\big[\ddot{\Omega}(0)-2\dot{\Omega}^2(0)\big]
-\frac{s^3}{6}\big[\dddot{\Omega}(0) -
6\ddot{\Omega}(0)\dot{\Omega}(0) + 6 \dot{\Omega}^3(0) \big] \bigg)
\bigg\} \nn\\
&=\Omega(s) \big\{ \delta\hat M(s) +
[\text{quadratic subtraction~polynomial}] \big\} \,.
\end{align}
The first part acts as the subtraction function in the dispersion
integral, including the singular term in $\kappa^2(s)$ $\propto
1/s$. The remainder---the subtraction terms obtained from
derivatives of the Omn\`es function at zero---can be discarded based
on arguments on the high-energy behavior in analogy to Appendix~B of
Ref.~\cite{Kang}.

\section{Flatt\'e parametrization } \label{Appendix.B}

In this Appendix, we briefly illustrate the effect of close-by
thresholds on the apparent width of a resonance signal. To be
specific, we will concentrate on the $Z_b(10610)$; yet, the
discussion applies in general for any structure located very close
to a strongly coupled threshold. Thus, a similar argument can also
be used for the $Z_b(10650)$. The discussion is based on the
Flatt\'e parametrization~\cite{Flatte} and is not new. It has been
emphasized in the case of the $f_0(980)$~\cite{Zou1993} (for
discussions of the $f_0(980)/a_0(980)$ states using the Flatt\'e
formalism, see also Ref.~\cite{Haidenbauer2005}).

In addition to $B\bar B^*$, the $Z_b(10610)$ has several two-body decay
channels such as $\Upsilon(nS)\pi$, $h_b(mP)\pi$, as well as the so-far unobserved
$\eta_b\rho$. All these bottomonium channels have thresholds much lower than the
$B\bar B^*$ one, and thus the sum of their partial widths can be approximated by
a constant width $\Gamma_1$. Then the Flatt\'e parametrization for the $Z_b$
spectral function is proportional to~\cite{Flatte}
\begin{equation}\label{eq.flatte}
\frac1{\big|{s-m_{Z_{b1}}^2}+i
m_{Z_{b1}}\left[\Gamma_1+\Gamma_{B\bar B^*}(s)\right]\big|^2}\,,
\end{equation}
where
\begin{equation}
\Gamma_{B\bar B^*}(s)=\frac{g^2}{8\pi m_{Z_{b1}}^2} \left[k\,\theta(
\sqrt{s}-m_B-m_{B^*} ) + i\, \kappa\, \theta(-
\sqrt{s}+m_B+m_{B^*} )\right] \,,
\label{eq:kkappa}
\end{equation}
with $g$ the
coupling constant of the $Z_{b1} B\bar B^*$ vertex, $k$ the center-of-mass
momentum of the $B$ meson, and $\kappa=|k|$.
It is easy to see that for either $\sqrt{s}>m_B+m_{B^*}$ or
$\sqrt{s}<m_B+m_{B^*}$ the denominator in Eq.~\eqref{eq.flatte} becomes larger
when $g$ increases.
Therefore, if the pole is located very close to the $B\bar B^*$ threshold,
which should be the case for the $Z_b(10610)$, a coupling to $B\bar B^*$
makes the $Z_b$ spectral function narrower than $\Gamma_1$.
\begin{figure}
 \centering \includegraphics[width=0.55\textwidth]{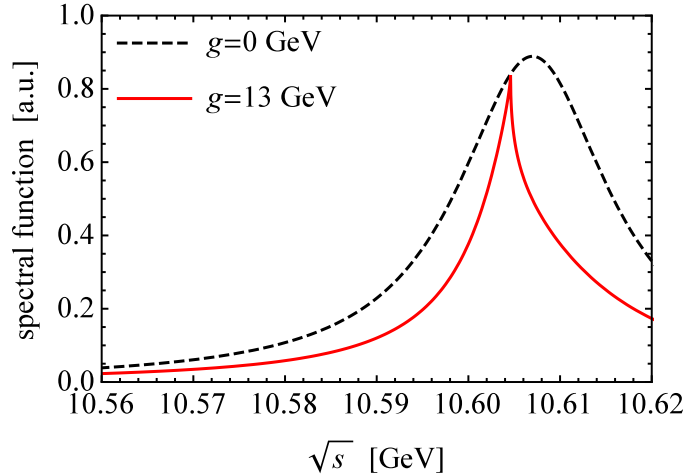}
 \caption{The spectral function of the $Z_b(10610)$ from Eq.~\eqref{eq.flatte}.
 Here, $m_{Z_{b1}}=10.607\GeV$ and $\Gamma_1=0.02\GeV$ are used for
 illustration.
 }
 \label{fig:flatte}
\end{figure}
This can be seen from Fig.~\ref{fig:flatte} where the spectral
function of the $Z_b(10610)$ is shown in arbitrary
units.\footnote{For $m_{Z_{b1}}=10.607\GeV$ and $\Gamma_1=0.02\GeV$,
the pole of Eq.~\eqref{eq.flatte} is located at
$(10.607-i\,0.01)\GeV$ for $g=0\GeV$ and $(10.600-i\,0.015)\GeV$ for
$g=13\GeV$.} Thus, we are led to conclude that $\Gamma_1$, the sum
of the partial decay widths other than that into $B\bar B^*$, in the
Flatt\'e parametrization should be larger than the nominal width of
the structure observed in the invariant mass distributions.

\end{document}